\documentclass[twocolumn,showkeys,amsmath,amssymb]{revtex4}

\usepackage{graphicx}
\usepackage{dcolumn}
\usepackage{bm}

\begin{document}

\preprint{APS-v0}

\title{Inverse Geometric Approach to the Simulation of the Circular Growth.\\
The Case of Multicellular Tumor Spheroids.}

\author{Branislav Brutovsky}
\email{bru@seneca.science.upjs.sk}
\author{Denis Horvath}
\email{horvath.denis@gmail.com}
\author{Vladimir Lisy}
\email{lisy@upjs.sk}
\affiliation{Institute of Physics, P.~J.~Safarik University, Jesenna 5, 04154 Kosice, Slovakia}
\date{\today}

\begin{abstract}
We demonstrate the power of the genetic algorithms to construct  
the cellular automata model simulating the growth of 2-dimensional
close-to-circular clusters revealing the desired properties, such as the growth rate
and, at the same time, the fractal behavior of their contours.
The possible application of the approach in the field of tumor modeling
is outlined.
\end{abstract}


\vspace*{6.5cm}

\maketitle

\section*{Introduction}

The surface phenomena, such as the growth, roughening and penetration into
the surrounding medium, as well as their modeling and efficient
control, represent long standing research topic in many branches of science and
technology. The special case is the close-to-circular growth observed
very often in biological systems, like bacterial colonies and tumors.
The growth models are usually classified into two groups,
(1) continuum, formulated through differential equations (DE),
and (2) discrete lattice models, most often represented by cellular automata (CA)
\cite{VonNeumann1966}, agent-based \cite{Mansury2002,Mansury2004b}, and Monte Carlo inspired models.
The CA models are typically constructed by {\it a priori}
(or microscopically reasoned) setting up the transition rules ({\it CA rules})
generating the desired behavior.
However, this {\it ad hoc} construction of the CA rules is,
due to high nonlinearity and multidimensionality of the CA models, 
often not appropriate in the case of more complicated behaviors. 
Below we propose the application of genetic algorithms
(GA) \cite{Holland1975} to retrieve the CA rules generating the growth of the clusters
revealing the desired features, such as the growth rate and fractal dimension
of the contour. 
In the here presented inverse geometric approach we quantify the differences
between the properties of a CA cluster growth and the fixed desired properties
(i. e. the quality of the corresponding CA rules) by the definition of the
{\it objective function}, and use the optimization technique to set the CA
rules which minimize the objective function value.
We demonstrate that
the approach is, in general, efficient enough to retrieve the CA rules
generating the growth of the CA cluster with the desired properties.
At the end we outline its possible contribution to the cancer research.

\section*{Methods}

{\it Cellular automata}
\cite{Wolfram1983}
were originally introduced
by John von Neumann as a possible idealization of biological
systems. In the simple case they consist of a 2D
lattice of sites $i, j = 0, \ldots, L_{\rm D}$, in the states $s^{(t)}_{ij} \in \{0,1\}$,
where $t = 0,\ldots, \tau$,
is the discrete time and $L_{\rm D}$ the size of the lattice.
During the $\tau$ time steps they evolve obeying the set of local
transition rules (CA rules) $\sigma$, formally written
\begin{equation}
\begin{array}{ll}
s^{(t+1)}_{ij} = \sigma(& \!\!s^{(t)}_{i-1j-1}, s^{(t)}_{ij-1}, s^{(t)}_{i+1j-1},
s^{(t)}_{i-1j}, s^{(t)}_{ij},\\
& \!\!s^{(t)}_{i+1j}, s^{(t)}_{i-1j+1}, s^{(t)}_{ij+1}, s^{(t)}_{i+1j+1}),
\end{array}
\label{CAmapping}
\end{equation}
which defines the CA rules $\sigma$ as the mapping
\begin{equation}
\sigma: \,
\underbrace{\{0,1\}\times\{0,1\}\times\ldots\times \{0,1\}}_{9}  
\rightarrow \{0,1\}\,.
\label{Mapping}
\end{equation}
The two basic update strategies are typically adopted: i) synchronous, when
the $t+1$ states (\ref{CAmapping}) of all the CA sites are computed
one by one in advance and the updates themselves are performed afterwards;
and ii) asynchronous, when a randomly chosen site is immediately updated.
The CA evolution is represented by the respective
point in $2^9$-dimensional binary space enabling, in principle,
the enormous number of $2^{512}$ possible global behaviors, predestining
the CA to be an efficient simulation and modeling tool
\cite{Toffoli1987}.
Inherent nonlinearity of the CA models is, however, a double-edged
sword. On the one hand, it enables to model a broad variety of
behaviors, from trivial to complex, on the other hand it results
in difficulty with retrieving the transition rules generating the
desired global behavior.


No well-established universal technique exists to solve the problem,
and, despite sporadic promising applications of the genetic algorithms
to solve the task \cite{Richards1990,Mitchell1994,Jimenez-Morales2002},
one typically implements the CA by intuitive ({\it ad hoc} or microscopically
reasoned) construction of the transition rules, or reduces the dimensionality
of the CA search space by incorporating a symmetry into the transition rules. 
Well known are {\it totalistic} rules,
when only the number of the adjacent sites in the state 1 (or 0) affects
the future state
of the respective site. Despite the crucial simplification, they provide
very complex behavior (e. g. Conway's Game of Life).

{\it Genetic Algorithms}
\cite{Holland1975}
are general-purpose search and
optimization techniques based on the analogy with Darwinian evolution of biological
species. In this respect, the evolution of a population of individuals is viewed as
a search for an optimum (in general sense) genotype. The feedback comes as 
the interaction of the resulting phenotype with environment. Formalizing 
the basic evolutionary mechanisms, such as mutations, crossing-over and survival
of the fittest, the fundamental theorem of GA was derived
({\it schema theorem}) which shows that the evolution is actually
driven by multiplying and combining {\it good} (quantified by an appropriate
objective function), correlations of traits (also called schemata, or hyperplanes).
The remarkable feature resulting from the schema theorem is the
implicit parallelism stating that by evaluating a (large enough) population of 
individuals, one, at the same time, obtains the information about the quality
of many more correlations of the traits.

In the following we use the GA optimization to retrieve the CA rules
producing the growth of the 2D CA cluster by the required rate as well
as fractal behavior of the {\it contour}, which is here understood
as the set of all the CA sites in the state 1 with at least one neighbor
in the state 0.


At the beginning, the chain of concentric circles ({\it learning patterns})
with randomly deformed close-to-circular contours
$p^{(t)}_{ij} \in \{0,1\},\ i, j = 0, \ldots, L_{\rm D}$,
for $t = 0,\ldots, \tau$, were generated accordingly to
\begin{equation}
\label{patterns}
p^{(t)}_{ij} =
\left\{
\begin{array}{ll}
0\,, & \mbox{for $\sqrt{i^2+j^2}>R^{(t)}+1$},\\
0\,, & \mbox{or $1$ drawn with probability \small$\frac{1}{2}$}\\
     & \mbox{for $|\sqrt{i^2+j^2}-R^{(t)}|\leq 1$},\\
1\,, & \mbox{for $\sqrt{i^2+j^2}<R^{(t)}-1$},
\end{array}
\right.
\end{equation} 
with linearly increasing tumor radius
\begin{equation}
R^{(t)} = R^{(0)}+At,
\label{LinRad}
\end{equation}
where $R^{(0)}$ is the initial radius of the CA cluster and $A$ ($<1$)
the desired growth rate constant.

The first objective of the optimization task solved by the GA was
to retrieve the CA rules (\ref{Mapping}) providing the growth from
the initial pattern $\{s_{ij}^{(0)}\}$, generated by the same relation
as $\{p_{ij}^{(0)}\}$ (\ref{patterns}), through the sequence of 
$\{s_{ij}^{(t)}\}$ ({\ref{CAmapping}), $t = 1,\ldots,\tau$, with the minimum mismatch
from the learning patterns $\{p_{ij}^{(t)}\}$ in the respective $t$,
quantified by the objective function
\begin{equation}
\label{objective1}
f_1(\sigma)  \equiv  \sqrt{\frac{1}{\tau}\sum^{\tau}_{t=1}
\left(\frac{\sum^{L_{\rm D}}_{i,j} p^{(t)}_{ij}+ 
\sum^{L_{\rm D}}_{i,j} s^{(t)}_{ij}}{1 + w_0 \sum^{L_{\rm D}}_{i,j} 
p^{(t)}_{ij}\delta_{p^{(t)}_{ij} s^{(t)}_{ij}}} 
\right)^2}\,,
\end{equation}
where $\delta$ is the standard Kronecker delta symbol, 
$w_0$ is the weight factor, in our case $w_0 = 2$.
The above expression of the objective function (\ref{objective1})
reflects the programming issues. The larger overlap 
of $\{s_{ij}^{(t)}\}$ with $\{p_{ij}^{(t)}\}$ enhances the 
denominator of Eq.~(\ref{objective1}), the prefactor $p_{ij}$ 
in the term $p_{ij} \delta_{p_{ij} s_{ij}}$ reduces
the computational overhead by ignoring the calcul
$\delta_{p_{ij} s_{ij}}$ for $p_{ij}=0$.

The second objective to the desired growth relates to the geometric properties
of the contour. Broadly accepted invariant measure expressing the contour
irregularity is the fractal dimension, $D_{\rm F}$.
Using the box-counting method
it can be calculated as
\begin{equation}
\label{FractalDimension}
D_{\rm F} = 
\lim_{\epsilon\to 0}\frac{\log M_{\rm B}(\epsilon)}{\log(1/\epsilon)},
\end{equation}
where $M_{\rm B}(\epsilon)$ is the minimum number of boxes of size $\epsilon$
required to cover the contour. Here, it has been determined as the slope in the
log-log plot of $M_{\rm B}(\epsilon)$ against $1/\epsilon$ using the standard
linear fit.
To obtain the CA rules generating
the cluster with the desired fractal dimension, $D_{\rm F}$,
the objective function (\ref{objective1})
has been multiplied by the factor
\begin{equation}
f_2(\sigma) = 1 + w_1 (D^{\tau}_{\rm F}-D_{\rm F})^2,
\label{objective2}
\end{equation}
where $D^{\tau}_{\rm F}$ is the fractal dimension of the cluster kept
after the $\tau $ steps with the CA rules $\sigma$, $w_1$ being the
weight.

Finally, the objective function is written
\begin{equation}
\label{objective}
f(\sigma) = f_1(\sigma) f_2(\sigma)\,, 
\end{equation}
and the optimum CA rules, $\sigma^{\ast}$, meet the requirement
\begin{equation}
f(\sigma^{\ast}) = \min_\sigma f(\sigma)\,.
\end{equation}

\section*{Results}

If not stated otherwise, all the below CA runs started from
the pattern $\{p_{ij}^{(0)}\}$ (\ref{patterns}), with the radius
$R^{(0)} = 5$. The size of the 2D CA lattice $L_{\rm D}$ varied
form 120 to 200 (reflecting the growth constant), with the periodic
boundaries, and the length of the CA evolution $\tau = 200$ steps.
The asynchronous update has been adopted.
The canonical GA search has been applied to retrieve
the set of the optimum CA rules, $\sigma^\ast$, 
which gives minimum objective function
values (\ref{objective1}, or \ref{objective}, respectively).
The size of the population was kept constant (1000 individui),
the probability of bit-flip mutation 0.001,
and the crossing over probability 1.
The ranking selection scheme was applied.
The length of the optimization was $3000$ generations.
\begin{figure}[h]
\includegraphics[scale=0.75]{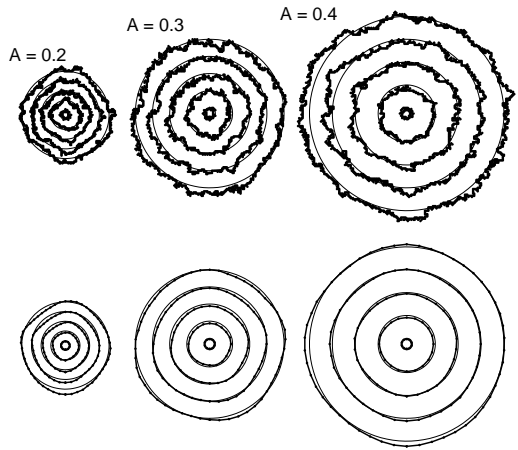}
\caption[]{
The comparison of the typical contours created by the CA rules
minimizing the criterion (\ref{objective1}) for the three different
required rates (mediated by the rate constant $A=0.2, 0.3, 0.4$ in the Eq.~(\ref{LinRad}))
(up, bold line) against the prescribed horizons after $t=0,50,100,150,200$
CA steps. Below are the respective average contours calculated from
100 repetitions of the growths.
The criterion (\ref{objective2}) has not been applied ($w_1 = 0$).
}
\label{growth512}
\end{figure}
Figs.~\ref{growth512}~to~\ref{w2} show the capability
of the above approach to fulfill the required objectives.
In Fig.~\ref{growth512} we present the comparison of the typical
growths created by the CA rules minimizing the criterion
(\ref{objective1}) with the respective prescribed rates (\ref{LinRad})
as well as the respective average contours
calculated from 100 repetitions for each of the growths. One can
see that in this case the CA rules preserve circular symmetry
for all the tested growth rates.
\begin{figure}[h]
\includegraphics[scale=0.9]{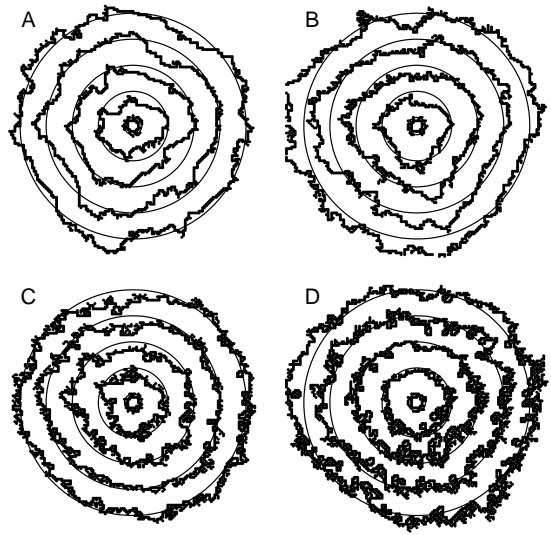}
\caption[]{
The comparison of the typical contours created by the CA rules
minimizing the criterion (\ref{objective}) for the four different
required fractal dimensions ($D_{\rm F}=1.1, 1.2, 1.3, 1.4$, denoted as
A, B, C, D) (up, bold line) with the prescribed horizons after $t=0,50,100,150,200$
CA steps. In Fig.~\ref{adf512} are the respective average contours calculated
from 100 repetitions.
The criterion (\ref{objective2}) has been applied with the weight $w_1 = 10$.
The rate constant $A$ in Eq.~(\ref{LinRad}) has been kept 0.3 in
all the cases.
}
\label{fd512}
\end{figure}
Fig.~\ref{fd512} shows the growing contours created by the CA rules
minimizing both the criteria, the desired rate and fractal dimension
of the contour, as implemented in (\ref{objective}).
The respective average contours (Fig.~\ref{adf512})
shows slight deviations from the circular symmetry.
\begin{figure}[h]
\includegraphics[scale=0.9]{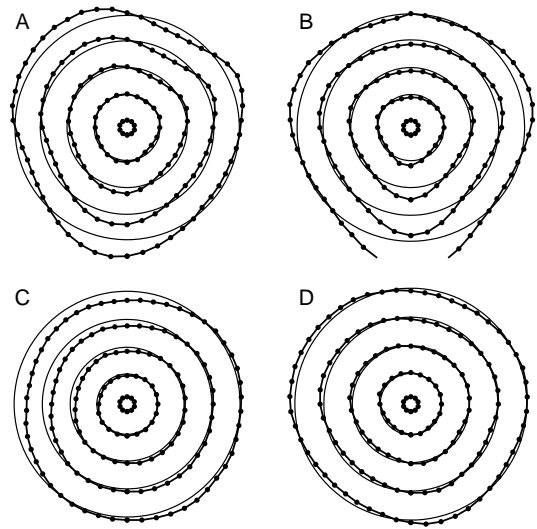}
\caption[]{
The average contours calculated from 100 repetitions for each of the
 A-D growth in Fig.~\ref{fd512}. The deviation from the circular
symmetry is obvious, especially if the lower fractal dimensions are
required during the optimization. 
}
\label{adf512}
\end{figure}
The above results demonstrate that the GA optimization procedure
can be used to retrieve the CA rules generating the close-to-circular growths
by desired rates and, at the same time, fractal dimensions
of their the contours (Fig.~\ref{w2}).

\clearpage

\begin{figure}[h]
\includegraphics[scale=0.42]{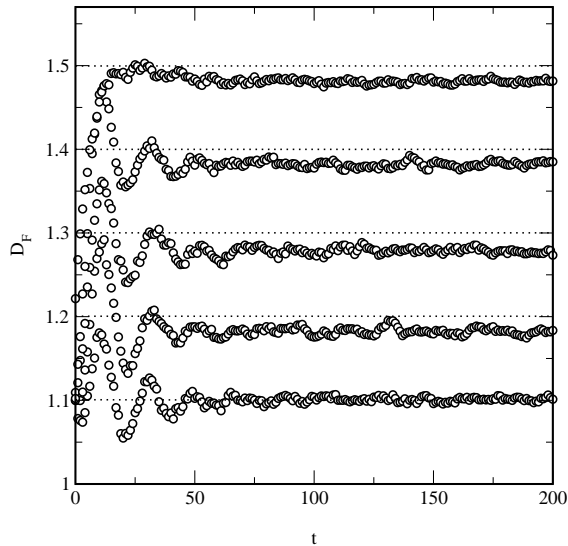}

\vspace{-2mm}

\caption[]{
Comparison of the average fractal dimensions of the contours, $D_{\rm F}$,
produced by the CA rules optimized to minimize the objective function
(\ref{objective}) against the respective desired fractal dimension
($D_{\rm F} = 1.1, 1.2, 1.3, 1.4, \ {\rm and} \ 1.5$).
The circles show the fractal dimensions for the respective CA rules
and time step. In all the cases $A=0.3$ and $w_1=10$ (Eqs.~\ref{LinRad} and
\ref{objective2}).
}
\label{w2}
\end{figure}

\vspace{-8mm}

\section*{Application to tumor growth modeling}

Here we present the application of the above approach in the field
of tumor modeling, enabling to study the tumor growth on the spatio-temporal
scales where the microscopic details have vanished and eventual universal
features of the growth emerge.
To understand the universal properties of the growth,
physicists have developed the {\it fractal and scaling analysis}
\cite{Barabasi1995}, enabling them to classify many growth processes
of seemingly different nature into the {\it universality classes} \cite{Odor2004}.
Led by similar motivations, the fractal and scaling methodology seems
to be potentially promising approach to study various aspects of carcinogenesis.
The morphology of growing tumor surface is the product of the interactions
of many biochemical agents, physical processes and geometrical constraints.
Pathologists have used the morphology and shapes of neoplasms to distinguish
between the malignant and benign forms as a rule of thumb.
Recently, a few papers
\cite{Landini1996,Spillman2004,Grizzi2005,Kikuchi2006,Risser2007}
have indicated
that the fractal dimension of the tumor contour, morphology or vascular
networks can be applied to distinguish between the normal and malignant tissue.
{\it In vitro} grown tumors usually form 2 (or 3) dimensional close-to-spherical
aggregations, called multicellular tumor spheroids (MTS) \cite{Delsanto2005}.
They can be viewed as the prototype of some real types of human tumors, such as
melanomas or gliomas. Allowing the strict control of nutritional and mechanical conditions,
the MTS are excellent experimental objects to test the validity
of the models of tumor growth.

The below outlined application of the inverse geometric approach
to tumor growth modeling is based on the morphometric MTS data by Bru
et al. \cite{Bru1998,Bru2003}, namely therein suggested linear increase
of the MTS mean radius, proliferation activity located at the tumor border
and the range of the fractal dimensions of the {\it in vitro} grown cell
colony contours between 1.1 and 1.4, such as HT-29 cell line (tumor type
colon adenocarcinoma, $D_{\rm F}=1.13$),
HeLa (cervix carcinoma, 1.25), Saos-2 (osteosarcoma, 1.34), etc.
Using the optimized CA models, we reproduce and interpret some of the features,
like the growth rate, response of the tumor to chemotherapy and scaling properties
of the contour.
We emphasize that in the below application the CA site does not
correspond to the only biological cell. We interpret it
as a 2D constant size window occupied by the above (or under) critical number
of tumor cells, corresponding to 2 states - 1 and 0, respectively.
Switching the state of the CA site from 0 to 1 (as well as from 1 to 0) during
the simulation simply means that the number of tumor cells in that
site has just crossed the critical value. The mechanism itself
stays out of the scope of this phenomenological approach, nevertheless
it encompasses the mean effect of proliferating activity, cells
migration, apoptosis, immune response, etc. 

\subsection*{Radius increase}

The simplest mathematical models of the tumor growth assume
an exponential increase of the tumor mass \cite{Shackney1993}
(which is however plausible for the early stage of carcinogenesis). 
Bru et al. \cite{Bru1998,Bru2003} have recently experimentally shown that
the MTS mass, $M^{(t)}$, actually grows linearly increasing its mean radius,
\begin{equation}
M^{(t)} = \pi(R^{(0)}+At)^2,
\label{LinearGrowth}
\end{equation}
where $R^{(0)}$ and $A$ retains their meanings (\ref{LinRad}).  

Fig.~\ref{radius} shows the log-log time dependence of the radius
of the two growing CA clusters applying the different rules,
optimized to produce the clusters with the fractal dimension
of their contours, $D_{\rm F}$, equal $1.1$, and $1.5$, respectively.
Their slopes differ negligible, nevertheless both are slightly
steeper than the slope of the idealized growth accordingly to
Eq.~(\ref{LinearGrowth}).

\begin{figure}[h]
\includegraphics[scale=0.42]{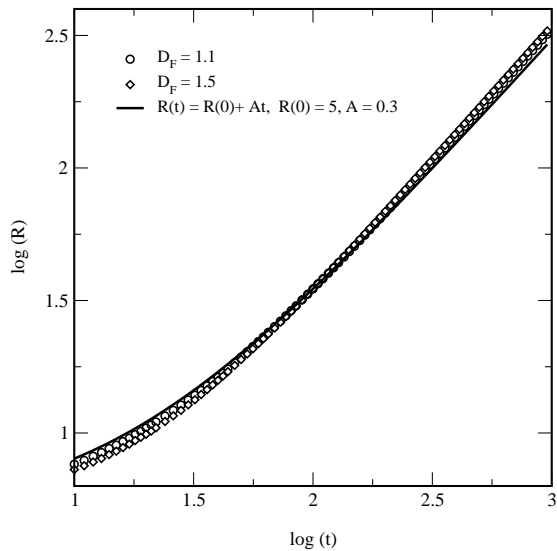}
\caption[]{
Comparison of the growing CA cluster radius created by the two CA
rules producing the contours with the different fractal dimensions $D_{\rm F}
= 1.1$ and $1.5$, respectively, against the required case
(Eq.~\ref{LinearGrowth}). In both the cases the rate constant $A = 0.3$.
}
\label{radius}
\end{figure}

Intuitively, the assumption of the linear increase of the radius
is consistent with the observation that proliferation activity is
located at the tumor contour. Nevertheless, the results (Fig.~\ref{radius})
show slight deviations from the linear increase of the radius, which we
attribute to longer than ideally circular contour of the cluster,
giving higher chance to proliferate.
On the other hand, the fractal dimension of the contour, $D_{\rm F}$,
corresponding to the roughness of the cluster contour, does not
influence the growth rate in a simple way. We have observed that
the CA cluster enforced to grow with the fractal dimension of the
contour close to 1 very often "freeze" repeating some simple
pattern which leads to significant deviations from circularity
\cite{Wolfram1985}.
As the deviations from the circularity of the tumor contour is
the obvious experimental fact as well, we find the hypothesis
of linear increase of the cluster mean radius valid for the spatio-temporal
scales where the deviations of the contour points from circularity
is negligible in comparison to the mean radius.

\subsection*{Simulation of chemotherapy}

To outline the possible therapeutic perspective of the above
inverse geometric approach we present the simulation of the impact
of chemotherapy to MTS growth.
Most chemotherapeutic drugs effectively damage fast proliferating
cells (such as tumor cells).
Here, the chemotherapy is included into the model in a similar way as in DE models
\cite{DeVladar2004} as the term inhibiting the production of tumor cells,
regardless of the microscopic background of the process.
For our purposes, the chemotherapy is viewed as any mechanism
(additional to those already imprinted on the CA rules by the optimization
procedure) converting the CA site from 1 (i. e. occupied by the above
critical number of tumor cells) to 0 (occupied by the under critical number
of tumor cells) with non-zero probability, represented by the free parameter,
called {\it efficiency of chemotherapy} and denoted as $c$}.
In our CA simulations the above mechanism
was realized simply by switching a randomly chosen CA site (of the contour
only - by definition in the state 1) to 0 with the probability $c$,
{\it instead} of applying the respective CA rule.

\begin{figure}[h]
\includegraphics[scale=0.66]{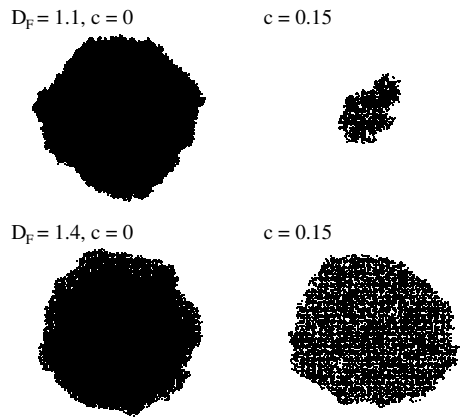}

\vspace{-3mm}

\caption[]{
Simulation of the impact of chemotherapy to the CA cluster after 200
time steps of the growth. The left column shows the CA clusters grown
accordingly to the CA rules optimized to generate the clusters with
the fractal dimension of their contours, $D_{\rm F}$, equal to 1.1 (up),
and 1.4 (bottom), respectively, without chemotherapy.
The right column depicts the cases received with the respective rules,
with the chemotherapy added ($c=0.15$).

\vspace{-3mm}

}
\label{chemotherapy}
\end{figure}
\begin{figure}[h]
\includegraphics[scale=0.4]{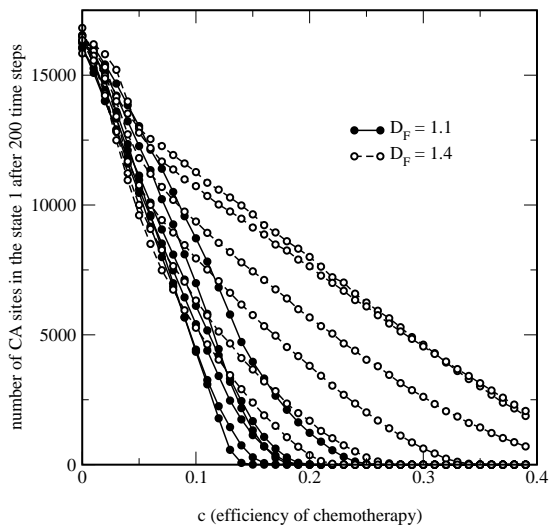}

\vspace{-3mm}

\caption[]{
Dependence of the impact of chemotherapy to the evolution of the CA clusters
generated by the two groups of the CA rules. The first set consists of 6 different
CA rules producing (in the absence of chemotherapy) the clusters with the fractal
dimension of their contours 1.1, the latter of 6 CA rules producing
the clusters with the fractal dimension 1.4. The CA rules
from the latter group produce the CA cluster growth less sensitive
to the chemotherapy.
}
\label{efficiency_chemotherapy}
\end{figure}

Fig.~\ref{chemotherapy} shows the typical impact of the simulated
chemotherapy on the growth of the CA clusters. One CA cluster was
grown with the transition rules which had been optimized to generate
smoother contours (quantified by its fractal dimension, $D_{\rm F} = 1.1$
, upper line), one CA cluster grown with
the rules optimized to generate more rough contours ($D_{\rm F} = 1.4$,
bottom line).
In the left column we compare the sizes and shapes of
both the clusters after 200 time steps grown 
in the absence of chemotherapy ($c = 0$), against
the respective cases with the chemotherapy added ($c = 0.15$, right column).
Fig.~\ref{efficiency_chemotherapy} shows the dependence of the CA cluster
growth on the efficiency of the chemotherapy for the two sets of transition
rules (generating the contours with $D_{\rm F} = 1.1$, and $1.4$,
respectively) in general.
The simulation results clearly demonstrate that the impact
of chemotherapy is more inhibiting to the CA clusters grown with
smoother contours than those growing with more rough contours.

Hopefully, the above demonstrated results could motivate
researchers to search for more quantitative relation 
between the fractal dimension of the tumor's contour
and the efficiency of chemotherapy.
Disposing with concrete morphometric tumor data, the above inverse
geometric approach could be used to build the corresponding
CA model to make predictions of the resistance to chemotherapy
or provide rough estimates of the therapeutic
efficiency for the respective {\it in vitro} grown tumor
cell lines.

\subsection*{Scaling behavior}

Inspired by the fractal and scaling concepts, Bru and coworkers
\cite{Bru1998,Bru2003} scrutinized the results of morphometric
experiments on a few tens different {\it in vitro} growing
tumors and cell colonies by the fractal and scaling analysis,
and suggested that the tumor reveals the basic features of Molecular
Beam Epitaxy (MBE) universality class \cite{Barabasi1995}
characterized by i)~a linear growth rate of the radius,
ii)~the proliferating activity at the outer border, and iii)~diffusion
at the tumor surface. Moreover, to characterize the tumor growth dynamics,
they determined the critical exponents \cite{Barabasi1995}, nevertheless
the issue is still under discussion \cite{Buceta2005,Bru2005}.

Here we outline the typical scaling behavior of the contours
of the growing CA clusters applying the CA rules 
obtained by the above optimization approach for the respective
desired growth rate and fractal dimension. We note that no
scaling properties have been explicitly included into the
optimization procedure.

A rough interface is usually characterized by the fluctuations 
of the contour points $r_{k,n}(t)$, where $k$ refers to the  
$k$th point of the $n$th segment of the arc length $l$ in time $t$.  
By defining the average
$\overline{r_n(t)} = (1/N_n(t))  
\sum_{k=1}^{N_n(t)} r_{k,n}(t)$, 
the {\it local interface width} is given by
\begin{equation}
\label{InterfaceWidth}
w(l,t) = \sqrt{\left<\frac{1}{N_n(t)}
\sum_{k=1}^{N_n(t)}
\left[ \,r_{k,n}(t)-  
\overline{r_n(t)} \,
\right]^2\,\right>}\,,
\end{equation}
where $N_1(t), N_2(t), \,\ldots\,,\, N_n(t), \,\ldots$ 
are the numbers of contour points 
along the $n$th angular segment 
of the arc length $l$.
The brackets $\langle \rangle$ mean 
the average over the different realizations
of the arc length $l$.

In general, the growth processes are expected to follow 
the Family-Vicsek ansatz
\cite{Family1985},
\begin{equation}
\label{FamilyVicsek}
w(l,t) = t^\beta \Phi(l/\xi(t))\,,
\end{equation}
with the asymptotic behavior of the scaling function
\begin{equation}
\label{guscal1}
\Phi(u) =
\left\{ 
\begin{array}{lll}
u^{\alpha}           &   \mbox{if}  &    u \ll  1\\
\mbox{const}         &   \mbox{if}  &    u \gg  1, 
\end{array}
\right.
\end{equation}
where $\alpha$ is the roughness exponent and characterizes
the stationary regime in which the height-height correlation length 
$\xi(t)\sim t^{1/z}$ ($z$ is so called dynamic exponent)
has reached a value larger than $l$. The ratio $\beta=\alpha/z$
is called the growth exponent and characterizes the short-time
behavior of the interface.

\begin{figure}[h]
\includegraphics[scale=0.42]{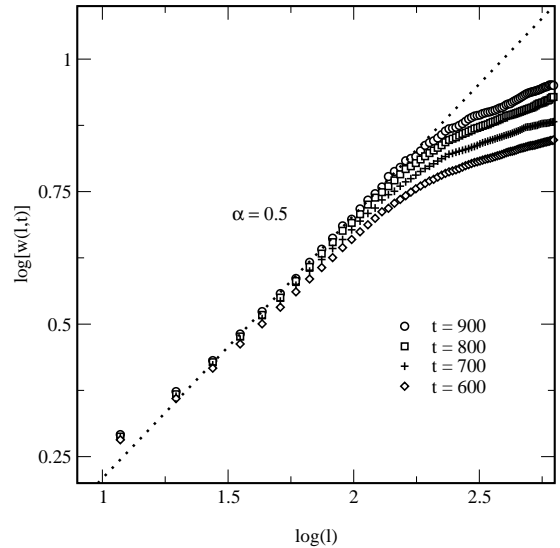}

\vspace{-2mm}

\caption[]{Log-log plot of $w(l,t)$. The region with
power law behavior corresponding to the exponent $\alpha$
shows the slope $0.5$.
}
\label{alfafit}
\end{figure}

\begin{figure}[h]
\includegraphics[scale=0.42]{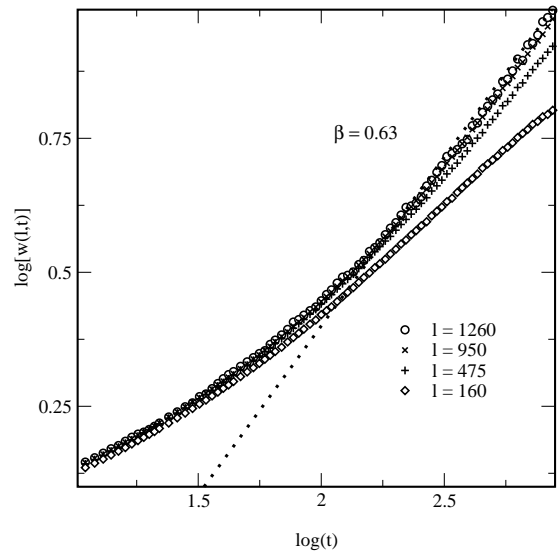}

\vspace{-2mm}

\caption[]{
Log-log plot of $w(l,t)$. The region with
power law behavior corresponding to the exponent $\beta$
shows the slope $0.63$.
}
\label{betafit}
\end{figure}

\begin{figure}[h]
\includegraphics[scale=0.42]{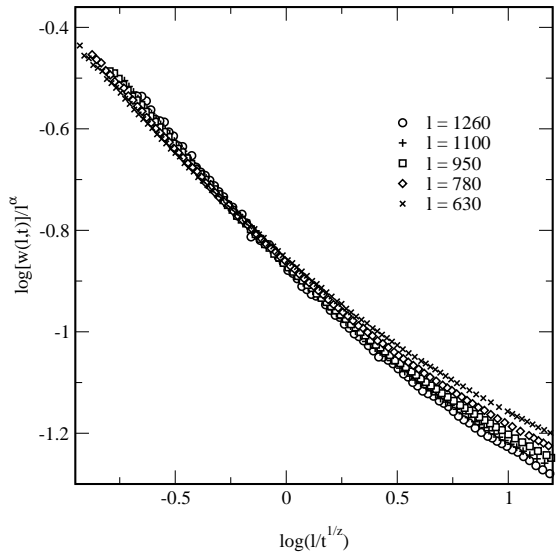}

\vspace{-2mm}

\caption[]{Scaling of $w(l,t)$ using the $\alpha$ and $z = 0.79$ exponents as provided
by the linear fits (Figs.~\ref{alfafit}~and~\ref{betafit}). Collapse of the curves for large $t$ is
obvious.
}
\label{collapse}
\end{figure}

Figs.~\ref{alfafit}~to~\ref{collapse} demonstrate the typical
scaling behavior of the growing CA cluster contours. By fitting the slope 
of the $w(l,t)$ in the respective regions (\ref{guscal1}) of the log-log
plot we have determined  the values of $\alpha$ and $\beta$ exponents,
giving the value of the exponent $z = 0.79$.

\vspace{3mm}

\vspace{3mm}

Performing the scaling analysis of 10 CA clusters growing accordingly
to the respective CA rules optimized to fulfill different combinations
of the desired rate constants, fractal dimensions and weights $w_1$
we have observed weak dependence of the exponent $z$ on the
growth rate (Tab.~\ref{exponents}).

\vspace{3mm}

\begin{table}[!h]
\begin{center}
\begin{tabular}{| c | c |c |c|}
\hline
&$A=0.2$	&$A=0.3$ 	&$A=0.4$\\
\hline
$\alpha$	&0.5 - 0.6	&0.41 - 0.56	&0.38 - 0.53\\
$\beta$		&0.63 - 0.76	&0.6 - 0.75	&0.68 - 0.89\\
$z$		&0.79		&0.63 - 0.74	&0.55 - 0.6\\
\hline
\end{tabular}
\end{center}
\caption[]
{
The dependence of the scaling exponents values on the growth rate of the
CA cluster growth.
}
\label{exponents}
\end{table}

The values of the scaling exponents differ from that presented in \cite{Bru2003}.
Nevertheless, there performed scaling analysis of the circular contours,
as well as the scaling analysis of growing circular contours in general,
is under discussion \cite{Bru2003,Buceta2005,Bru2005}.

\section*{Conclusion}

During the recent years cancer research has developed into a very
active scientific field with many concepts from other scientific
areas, such as statistical physics, applied mathematics,
nonlinear science and modeling
\cite{Baish2000,Delsanto2000,Deisboeck2001,Ferreira2002,Chignola2000,Galam2001,Gazit1995,Scalerandi2002,Sole2003}
penetrating into it.
The ultimate aim of the modeling is, in general, to provide correct
predictions for the real systems. These traditionally come from
the deep knowledge of the problem (microscopic models, {\it ab initio}
approaches) or from phenomenology (e. g. represented by
the growth laws written in the form of the first order differential equations).
The microscopic models of tumor growth are typically based on
the conceptions such as cell migration, proliferation, chemotaxis,
apoptosis, etc.
However, involving the plethora of parameters, these models often face
to conventional problems of complex systems - subjective choice of
the model parameters and their questionable measurement, both leading
to the loss of predictive power.
On the other hand, macroscopic models often miss adequacy \cite{Castorina2006}.
The way out of the obstacle could be based on the phenomenological
universalities involving phenomena like scale laws, complexity,
and nonlinearity \cite{Castorina2006}.
Here presented approach can be seen as the instance of that way
of thinking, benefiting from the demonstrated possibility to simulate
the MTS growth by the respective GA optimized CA model.
Implementing some of the features of the MTS growth into the CA model
one can eventually observe the feature(s) which has not originally been
included into the model and, consequently, try to
identify the feature(s) in the MTS growth itself.

To illustrate the future perspective of the above CA approach
in cancer research we propose the hypothesis of the existence
of the {\it saturation time} dividing the short time behavior
of the local interface width (\ref{InterfaceWidth})
from its universal, time-independent behavior, which is obvious
in the simulated CA clusters growth (Figs.~\ref{alfafit} and \ref{betafit}).
In general, during the roughening of a contour the lateral correlations
develop. At the beginning, different sites on the contour are correlated
only to the characteristic distance, the correlation length (\ref{FamilyVicsek}).
The local interface width increases as a power of time until the correlation length reaches
the size of the contour, the time is called {\it saturation time}.
After that all the sites become correlated and the local
interface width reaches a (time independent) saturation value, which
increases as a power of the system size.
However, if the system, in addition to roughening of the contour,
increases in size, the question whether
and when the saturation is reached (or lost, if started
as correlated) arises. Consequently, the following questions
emerge: Show the different tumors different saturation times (and values)
of the local interface width? Does the behavior of the local interface
width relates to malignity of tumors? Can it be purposefully affected
by the chemotherapy?
The finding of eventual relation between the malignity
(or proliferation activity) and the saturation time supported
by the experimental evidence could lead to the development
of the new therapeutic approach based on the (re)establishing
of the correlations.

The most salient feature of the above approach is that
neither microscopically based parameters in the traditional
sense, such as the cell migration, proliferation, chemotaxis, apoptosis,
etc, are required nor the relations between them are {\it a priori} built.
Both stay hidden in the transition rules, optimized to fulfill
chosen objectives, and acquire statistical (or even symbolic) meaning.
For example, the proliferation activity can be deduced only from
the increase (or decrease) of the CA cluster size, not from the local changes
of the states of individual CA sites, as on the here adopted scale
it is impossible to distinguish between the contributions
of cell proliferation, migration, apoptosis, etc.
On the other hand, some statistically modeled mechanisms
(such as the response to drugs) can be included in a quite
straightforward manner.

\newpage

{\it 
The authors acknowledge financial support from VEGA, Slovak Republic
(Grants No. 1/4021/07 and 1/3033/06).
High performance computing part of this work has been performed under
the Project HPC-EUROPA (RII3-CT-2003-506079), with the support
of the European Community - Research Infrastructure Action under
the FP6 "Structuring the European Research Area" Programme.
}


\end{document}